\documentclass[conference]{IEEEtran}
\IEEEoverridecommandlockouts
\makeatletter
\def\@IEEEauthorblockconfadjspace{-1.6em}
\makeatother
\usepackage[draft]{hyperref}
\usepackage{cite}
\usepackage{url}
\usepackage{amsmath,amssymb,amsfonts}
\usepackage{algorithmic}
\usepackage{graphicx}
\usepackage{textcomp}
\usepackage{xcolor}
\usepackage{booktabs}
\usepackage{multirow}
\usepackage{tikz}
\usepackage{pgfplots}
\usepackage{orcidlink}

\pgfplotsset{compat=1.17}
\usetikzlibrary{arrows.meta, positioning, shapes.geometric, fit, backgrounds}
\def\BibTeX{{\rm B\kern-.05em{\sc i\kern-.025em b}\kern-.08em
    T\kern-.1667em\lower.7ex\hbox{E}\kern-.125emX}}
\graphicspath{{./}{../output/}}

\begin{document}
\bstctlcite{IEEEexample:BSTcontrol}

\title{Layer-wise Cross-Lingual Depression Detection from Speech: Analysis with Contrastive Alignment}

\author{
  \IEEEauthorblockN{
    Anisha Pattanayak\scalebox{0.75}{\orcidlink{0009-0005-2556-4472}}$^{\star}$,
    Hanie Kang\scalebox{0.75}{\orcidlink{0009-0005-8023-9672}}$^{\dagger}$,
    Huang-Cheng Chou\scalebox{0.75}{\orcidlink{0000-0003-2125-5689}}$^{\star\ddagger}$,
    Sudarsana Reddy Kadiri\scalebox{0.75}{\orcidlink{0000-0001-5806-3053}}$^{\star\ddagger}$,
    Shrikanth Narayanan\scalebox{0.75}{\orcidlink{0000-0002-1052-6204}}$^{\star\dagger\ddagger}$
  }\\
  \IEEEauthorblockA{
    $^{\star}$\textit{Ming Hsieh Department of Electrical and Computer Engineering, University of Southern California, USA}\\
    $^{\dagger}$\textit{Thomas Lord Department of Computer Science, University of Southern California, USA}\\
    $^{\ddagger}$\textit{Signal Analysis and Interpretation Laboratory (SAIL), University of Southern California, USA}
  }
}

\maketitle
\begin{abstract}
Significant disparities exist in the diagnosis and clinical presentation of depression across different linguistic populations.
Speech-based depression detection performs well monolingually, but cross-lingual generalization remains an open challenge.
A key reason is that prior work uses segment-level random splits without speaker grouping, leading to identity leakage that inflates reported metrics.
We propose CLeaD, a supervised contrastive alignment framework that maps WavLM embeddings from English and Mandarin into a shared clinical space, without parallel data or target-language fine-tuning.
Evaluating 52 Mandarin speakers, contrastive alignment modestly outperforms the baseline (F1: 0.640 vs. 0.622) under leave-one-speaker-out evaluation.
It also improves depressed-class recall at intermediate layers (7--8), though the small test set limits generalizability.
Two findings remain robust: model scaling degrades cross-lingual performance while improving monolingual English, and a controlled within-pipeline ablation shows that speaker-identity leakage alone inflates Mandarin F1 from 0.628 to 0.856, an artifact we reproduce and quantify.
\end{abstract}
\vspace{-1.6mm}
\section{Introduction}
Depression is a common psychiatric disorder, affecting approximately 322 million people worldwide in 2023 and substantially increasing disability and suicide risk \cite{who2023depression}.
In clinical practice, diagnosis relies on structured interviews and validated rating scales such as the Patient Health Questionnaire (PHQ-8/9).
These instruments require trained professionals to administer, limiting deployment in low-resource healthcare settings \cite{cummins2015review}.
Speech has emerged as a promising non-invasive signal for automated screening.
Depression measurably alters acoustic characteristics, including reduced pitch variability, slower speaking rate, increased pause frequency, and lower vocal energy \cite{Arevian2019Clinicalstatetrackingin,kappen2023speech}.
Self-Supervised Learning (SSL) models like HuBERT \cite{hsu2021hubert} and WavLM \cite{chen2022wavlm} achieve strong results.
Studies show their intermediate layers capture affective and prosodic cues better than the outer layers \cite{maji2024investigation, han2024crosslingual}.
A critical gap in cross-lingual transfer remains.
Most systems are English-centric and transfer poorly to other languages.
Mandarin Chinese is especially challenging, as its lexical pitch contours can mask the prosodic depression markers relied upon by English models \cite{wang2015pitch, kirmayer2001cultural}.
To address this gap, we propose CLeaD.
This paper makes four contributions:
\begin{itemize}
    \item \textbf{Proposed Framework:} We introduce CLeaD, a supervised contrastive alignment framework \cite{khosla2020supervised} that maps English and Mandarin WavLM embeddings into a shared clinical space without requiring parallel data or target-language fine-tuning.
    \item \textbf{Layer-wise Transfer Analysis:} We conduct a systematic layer-wise representation analysis of WavLM-Base-Plus and WavLM-Large across six transfer conditions, validated by 95\% bootstrap confidence intervals with 2000 resamples.
    \item \textbf{Quantification of Evaluation Artifacts:} We quantify speaker-identity leakage inside our own pipeline, showing that segment-level random splits inflate Mandarin F1 from 0.628 to 0.856 while feature-scaling leakage contributes less than 0.001.
    \item \textbf{Rigorous Clinical Evaluation:} We validate our framework through a leave-one-speaker-out (LOSO) evaluation on 52 MODMA speakers, ablation studies (e.g., CLeaD w/o SupCon), and hyperparameter sensitivity analysis. Furthermore, we prioritize depressed-class speaker recall to align with clinical screening priorities, alongside standard LOSO speaker-level F1 metrics.
\end{itemize}

\section{Related Work}
\subsection{Depression Detection from Speech}
Early approaches relied on hand-crafted features such as MFCCs, pitch statistics, and shimmer paired with SVMs \cite{cummins2015review, albuquerque2021association}.
Recurrent architectures improved temporal modeling \cite{alhanai2018detecting}.
SSL models have since outperformed hand-crafted pipelines, with HuBERT and wav2vec~2.0 \cite{baevski2020wav2vec} offering better generalization \cite{zhang2021depa, pepino2021emotion}.

\subsection{Layer-Wise Analysis of SSL Models}
Maji et al.~\cite{maji2024investigation} showed via layer-wise analysis of HuBERT that intermediate layers work best for depression detection, monolingually and cross-lingually, and Han et al.\ \cite{han2024crosslingual} reached analogous conclusions for cross-lingual emotion recognition.
WavLM \cite{chen2022wavlm} extends HuBERT with a masked denoising objective, improving robustness for variable-quality clinical recordings, while ML-SUPERB \cite{shi2023mlsuperb} showed base SSL models generalize better cross-lingually than large ones, which our work confirms for clinical depression detection.
Li et al.\ \cite{li2025hierarchical} fuse several SSL layers on DAIC-WOZ and MODMA, indicating single-layer extraction leaves information unused, though their within-corpus evaluation motivates future work rather than a baseline here; Zhang et al.\ \cite{zhangEntangle} link representational entanglement to weak transfer, consistent with our scale reversal (Sec.~\ref{sec:scale}).
Relative to these studies, CLeaD combines a label-driven objective that \emph{actively reshapes} the embedding space for cross-lingual clinical transfer with a controlled within-pipeline quantification of speaker-identity leakage.

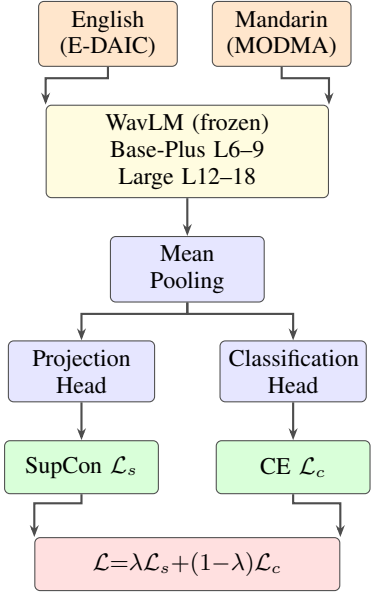
\begin{figure}[!t]
\centering
\begin{tikzpicture}[scale=0.72, transform shape,
  box/.style={rectangle, rounded corners=2pt, draw=black!70, fill=blue!10,
              text width=1.8cm, align=center, font=\small, minimum height=0.7cm},
  langbox/.style={rectangle, rounded corners=2pt, draw=black!70, fill=orange!20,
              text width=1.6cm, align=center, font=\small, minimum height=0.7cm},
  modelbox/.style={rectangle, rounded corners=2pt, draw=black!70, fill=yellow!15,
              text width=3.5cm, align=center, font=\small, minimum height=0.7cm},
  lossbox/.style={rectangle, rounded corners=2pt, draw=black!70, fill=green!15,
              text width=1.8cm, align=center, font=\small, minimum height=0.7cm},
  arr/.style={-{Stealth[length=4.5pt]}, thick, black!70}
]
\node[langbox] (en) {English\\(E-DAIC)};
\node[langbox, right=0.45cm of en] (zh) {Mandarin\\(MODMA)};
\node[modelbox, below=0.5cm of en, xshift=1.05cm] (bp)
  {WavLM (frozen)\\Base-Plus L6--9\\Large L12--18};
\node[box, below=0.5cm of bp] (pool) {Mean Pooling};
\node[box, below=0.5cm of pool, xshift=-1.4cm, text width=1.7cm] (proj)
  {Projection\\Head};
\node[box, below=0.5cm of pool, xshift=1.4cm, text width=1.75cm] (cls)
  {Classification\\Head};
\node[lossbox, below=0.5cm of proj] (supcon) {SupCon $\mathcal{L}_s$};
\node[lossbox, below=0.5cm of cls] (ce) {CE $\mathcal{L}_c$};
\node[lossbox, below=0.5cm of pool, yshift=-2.6cm, text width=3.8cm, fill=red!12]
  (joint) {$\mathcal{L}{=}\lambda\mathcal{L}_s{+}(1{-}\lambda)\mathcal{L}_c$};
\draw[arr] (en.south) -- ++(0,-0.15) -| (bp.north west);
\draw[arr] (zh.south) -- ++(0,-0.15) -| (bp.north east);
\draw[arr] (bp) -- (pool);
\draw[arr] (pool.south) -- ++(0,-0.15) -| (proj.north);
\draw[arr] (pool.south) -- ++(0,-0.15) -| (cls.north);
\draw[arr] (proj) -- (supcon);
\draw[arr] (cls) -- (ce);
\draw[arr] (supcon.south) -- ++(0,-0.12) -| (joint.north west);
\draw[arr] (ce.south) -- ++(0,-0.12) -| (joint.north east);
\end{tikzpicture}
\caption{CLeaD pipeline. Both languages share a frozen WavLM extractor. Setting $\lambda{=}0$ gives CLeaD w/o SupCon.}
\label{fig:pipeline}
\vspace{-6mm}
\end{figure}

\subsection{Cross-Lingual Depression Detection}
Cross-lingual depression detection predates SSL models: Kiss and Vicsi \cite{kiss2017mono} studied mono- and multi-lingual prediction from hand-crafted features, Mitra et al.\ \cite{mitra2015crosscorpus} showed cross-corpus prediction degrades sharply, and Qin et al.\ \cite{qin2025crosslinguistic} align feature covariances across languages (CORAL-family). CLeaD differs in \emph{what} it aligns: clinical label rather than the distribution statistics used by DANN, CORAL and MMD, via supervised contrastive learning on frozen WavLM with a 240-run layer study under LOSO; this is a design difference rather than a demonstrated gain, since we train no adaptation baseline.

Chen \cite{chen2025layerwise} is the closest study, applying layer-wise HuBERT analysis to English and Mandarin within the AVEC framework \cite{valstar2016avec}, but it evaluates on CMDC \cite{zou2023cmdc}, not MODMA, and its Mandarin F1 near 0.95 is monolingual, so it is not comparable with our cross-lingual numbers. Its pipeline splits segments randomly without speaker grouping, which we confirmed from the public codebase, and such splitting is widespread: Qin et al.\ \cite{qin2025crosslinguistic} split 60/20/20 at sample level on this pair, so their 81.1\% Chinese accuracy is likewise not comparable, and Danylenko and Unold \cite{danylenko2026pitfalls} find only five DAIC-WOZ studies meeting minimal reproducibility standards. 
That audit covers ungrouped splits only, not grouped work such as Maji et al.\ \cite{maji2024investigation}.

We therefore quantify the effect \textit{within our own pipeline}, on the same WavLM Base-Plus backbone and MODMA.
Table~\ref{tab:leakage} shows that speaker identity leakage alone inflates Mandarin LR F1 from 0.628 to 0.856 and AUC from 0.706 to 0.933, a 0.23 F1 jump from splitting segments rather than speakers, while feature scaling leakage adds under 0.001, isolating speaker identity as the primary driver of inflated metrics under ungrouped splits.

\vspace{-1mm}
\section{Methodology}
\vspace{-1mm}
\subsection{Datasets}
\vspace{-1mm}

\textit{E-DAIC (English).} The Extended DAIC-WOZ corpus \cite{gratch2014distress}, used in the AVEC challenge series \cite{ringeval2019avec}, consists of 219 semi-structured clinical interviews in English, conducted by a virtual interviewer.
PHQ-8 $\geq$ 10 defines depression labels.
E-DAIC provides an official challenge partition, but its test-set labels are withheld; we therefore re-partition the publicly labeled data into stratified, speaker-independent splits (Sec.~\ref{sec:split}), yielding 122 participants: 75 train (22 depressed/53 healthy), 24 validation (7/17), and 23 test (6/17).
\textit{MODMA (Mandarin).} MODMA \cite{cai2022modma} is a Mandarin clinical depression corpus with structured interviews and PHQ-9 labels.
MODMA is released without a canonical train/test partition, so we apply the same procedure (Sec.~\ref{sec:split}) to 52 participants: 31 train (13 depressed/18 healthy), 11 validation (5/6), and 10 test (5/5).
PHQ-8 is the eight-item Patient Health Questionnaire depression scale; PHQ-9 adds the suicidal-ideation item. We label a participant \emph{depressed} (probable major depressive disorder, MDD) when the total is $\geq 10$ and \emph{healthy control} (HC) otherwise; 10 is the standard screening cut-off on both instruments, following prior cross-corpus work \cite{cummins2015review}.
Demographic differences between the two corpora (age, recording conditions, interview structure) represent potential confounds for cross-lingual transfer beyond language alone.

\begin{table}[!t]
\caption{Within-pipeline leakage ablation (WavLM Base-Plus, Layer~6, Logistic Regression).}
\vspace{-5mm}
\label{tab:leakage}
\begin{center}
\begin{tabular}{lcccc}
\toprule
\textbf{Dataset} & \textbf{Airtight} & \textbf{+Scale} & \textbf{+Speaker} & \textbf{Full} \\
 & \textbf{F1/AUC} & \textbf{Leak} & \textbf{Leak} & \textbf{Leaky} \\
\midrule
E-DAIC (EN) & .481/.643 & .481/.643 & .747/.888 & .747/.888 \\
MODMA (ZH) & .628/.706 & .628/.706 & .856/.933 & .856/.933 \\
\bottomrule
\end{tabular}
\end{center}
\vspace{-1mm}
{\footnotesize Values are F1/AUC. Columns progressively relax the protocol: \textbf{Airtight} is fully speaker-independent; \textbf{+Scale Leak} additionally fits feature normalization on pooled train+test data; \textbf{+Speaker Leak} additionally lets segments from the same speaker span train and test; \textbf{Full Leaky} applies both. \par}
\vspace{-6mm}
\end{table}

\vspace{-1mm}
\subsection{Data Preprocessing}
\vspace{-1mm}
All audio is resampled to 16~kHz and converted to mono.
For E-DAIC, participant speech is isolated using transcript-aligned timestamps, excluding interviewer turns.
Both datasets are segmented into 3-second chunks with 50\% overlap, each inheriting the speaker's binary depression label. The 3-second window balances prosodic context against the number of training segments, while the 50\% overlap augments the limited clinical data; short overlapping windows of this kind are common practice in SSL-based speech classification \cite{maji2024investigation}.
\subsubsection{Speaker-independent splitting}\label{sec:split}
WavLM embeddings encode substantial speaker identity.
When segments from the same speaker appear in both training and test sets, a classifier can recognize the individual rather than depression-relevant patterns, which is speaker identity leakage.
We use stratified group-aware splits with participant ID as the grouping variable.
All segments from a given speaker are assigned to exactly one partition, with label stratification preserving class balance.
\subsubsection{Mixed-language training}
Because E-DAIC contains far more segments than MODMA, we randomly subsample the English segments to match the MODMA segment count. This balances the two languages within each training batch, preventing the majority language from dominating the loss and ensuring that cross-lingual positive pairs (one English and one Mandarin sample of the same clinical label) occur in every CLeaD batch.

\vspace{-1mm}
\subsection{WavLM Feature Extraction}
\vspace{-1mm}
WavLM-Base-Plus (12 layers, $d{=}768$) and WavLM-Large (24 layers, $d{=}1024$) are used as frozen extractors, both pretrained on 94K hours of English speech.
We extract hidden states from layers 6--9 of Base-Plus and layers 12, 14, 16, and 18 of Large. Because Large has twice as many layers, these ranges occupy the same \emph{relative} depth (roughly the middle of each stack), so comparing them isolates depth-in-proportion rather than absolute layer index. Mean pooling over $F$ frames at layer $l$ gives:
\begin{equation}
\bar{\mathbf{h}}^{(l)} = \frac{1}{F}\sum_{f=1}^{F}\mathbf{h}^{(l)}_f \in \mathbb{R}^{d},
\label{eq:pool}
\end{equation}
where $\mathbf{h}^{(l)}_f \in \mathbb{R}^{d}$ is the WavLM hidden state of frame $f$ at layer $l$, $F$ is the number of frames in the utterance, $d$ is the embedding dimension ($768$ for Base-Plus, $1024$ for Large), and $\bar{\mathbf{h}}^{(l)}$ is the resulting mean-pooled utterance-level embedding used as input to CLeaD.
Freezing WavLM ensures performance differences reflect layer representational properties, not fine-tuning artifacts. Fig.~\ref{fig:pipeline} shows the full pipeline.

\textit{Layer selection rationale.}
Layer 6 is used as the representative layer in comparative Tables~\ref{tab:leakage} and~\ref{tab:monolingual} for consistency with prior work \cite{maji2024investigation}, which reports Base-Plus Layer 6 as the strongest monolingual layer. CLeaD's peak F1 occurs at Layer 8 (0.561) with best recall at Layers 7--8 (4/5 Dep-Rec), as detailed in Table~\ref{tab:ablation}. These two choices are complementary rather than contradictory: Layer 6 anchors cross-paper comparison while the full ablation reveals the actual peak.

\vspace{-1.5mm}
\subsection{Contrastive Learning for Depression Detection (CLeaD)}
CLeaD is a two-head network trained on frozen WavLM embeddings. The \textit{projection head} maps $\mathbf{h}$ to a normalized 128-dim space:
\begin{equation}
\mathbf{z} = \ell_2\text{-norm}\!\left(\mathbf{W}_2\,\text{ReLU}(\text{BN}(\mathbf{W}_1\mathbf{h}))\right),
\label{eq:proj}
\end{equation}
where $\mathbf{h}\in\mathbb{R}^{d}$ is the pooled WavLM embedding from Eq.~\eqref{eq:pool}, $\mathbf{z}\in\mathbb{R}^{128}$ is the resulting $\ell_2$-normalized projection, $\mathbf{W}_1{\in}\mathbb{R}^{256\times d}$ and $\mathbf{W}_2{\in}\mathbb{R}^{128\times 256}$ are learnable weight matrices, ReLU$(\cdot)$ is the rectified linear unit, BN denotes batch normalization, and dropout (rate 0.3) is applied before $\mathbf{W}_2$. The \textit{classification head} maps $\mathbf{z}$ to class logits through a two-layer MLP with ReLU activation and dropout, mapping $\mathbb{R}^{128} \to \mathbb{R}^{64} \to \mathbb{R}^{2}$ with dropout rate 0.2 between layers. The training objective combines a supervised contrastive loss and a classification loss:
\begin{equation}
\mathcal{L} = \lambda\mathcal{L}_{\text{SupCon}} + (1-\lambda)\mathcal{L}_{\text{CE}},\quad\lambda=0.5,
\label{eq:loss}
\end{equation}
where $\lambda\in[0,1]$ balances the two terms (set to $0.5$), $\mathcal{L}_{\text{CE}}$ is the class-weighted cross-entropy on the classifier logits, and $\mathcal{L}_{\text{SupCon}}$ is the supervised contrastive loss:
\begin{equation}
\mathcal{L}^i_{\text{SupCon}} = \frac{-1}{|P(i)|}\sum_{p\in P(i)}\log\frac{\exp(\mathbf{z}_i{\cdot}\mathbf{z}_p/\tau)}{\sum_{a\in A(i)}\exp(\mathbf{z}_i{\cdot}\mathbf{z}_a/\tau)},\;\tau{=}0.1,
\label{eq:supcon}
\end{equation}
where $i$ indexes an anchor sample in the batch, $A(i)$ is the set of all other samples in the batch, $P(i)\subseteq A(i)$ is the subset sharing the anchor's clinical label, $\mathbf{z}_i,\mathbf{z}_p,\mathbf{z}_a$ are the corresponding projection embeddings, $(\cdot)$ denotes the dot product, and $\tau$ is the temperature. Because $P(i)$ includes same-label samples from \textit{both} languages, the loss pulls depressed embeddings from English and Mandarin together. Single-view contrastive learning is used; pitch-shifting and time-stretching are excluded to preserve depression-relevant prosodic markers. \textbf{CLeaD w/o SupCon} uses the same architecture but $\lambda{=}0$ (CE only), directly isolating what contrastive alignment adds over a more expressive nonlinear classifier.

\vspace{-1.5mm}
\subsection{Baselines and Evaluation}
Baselines on identical frozen embeddings: Logistic Regression (LR, matching Chen \cite{chen2025layerwise}), SVM-Linear, and bidirectional GRU sequence classifier. All baselines used their standard recommended settings (LR: balanced weighting, max 1000 iterations; SVM: balanced weighting; GRU: default architecture). No additional hyperparameter search was performed for baselines, which may mildly understate their ceilings; this is acknowledged as a limitation.
Six conditions: EN$\to$EN, ZH$\to$ZH (monolingual); EN$\to$ZH, ZH$\to$EN (zero-shot); MIX$\to$EN, MIX$\to$ZH (mixed). Unless otherwise noted, \textbf{F1} denotes the binary F1 of the depressed (positive) class, and \textbf{Spk~F1} its speaker-level counterpart; we report positive-class F1 because it directly reflects screening sensitivity. For Mandarin conditions, speaker-level majority voting gives a depressed-class recall count (\textbf{Dep-Rec}):
\begin{equation}
\hat{Y}_{\text{spk}} = \mathbf{1}\!\left[\textstyle\sum_{t=1}^{T}\hat{y}_t \geq T/2\right],
\label{eq:vote}
\end{equation}
where $\hat{y}_t\in\{0,1\}$ is the prediction for segment $t$, $T$ is the speaker's segment count, $\mathbf{1}[\cdot]$ is the indicator, and $\hat{Y}_{\text{spk}}$ is the speaker label. Consistent with the released implementation, a tie is assigned to the depressed class ($\sum_t\hat{y}_t=T/2$), a conservative choice for screening. \textbf{Dep-Rec} and \textbf{HC-Rec} count correctly identified depressed and healthy-control speakers, respectively, among the five per class in the held-out MODMA test set. Because each count changes in 20-point increments, these values in Tables~\ref{tab:crosslingual} and~\ref{tab:ablation} are descriptive rather than statistically definitive.
LOSO on all 52 MODMA speakers (Table~\ref{tab:loso}) is our primary Mandarin metric for statistical claims.
To assess statistical significance, we compute 95\% confidence intervals (CIs) using a non-parametric bootstrap approach with 2000 resamples.
Throughout this paper, these intervals are reported in brackets as \text{[lower bound, upper bound]} \cite{ferrer2022bootstrap}.

\vspace{-1mm}
\section{Experimental Setup}
\vspace{-1mm}
\subsection{Training Configurations}
\vspace{-1mm}
Monolingual conditions train and test on the same language, giving an upper bound without cross-lingual transfer; zero-shot conditions train on one language and test on the other with no adaptation, the hardest practical setting.
Mixed conditions combine both languages per batch, the only setting where the SupCon alignment objective can engage, which is why CLeaD's Dep-Rec gains are specific to MIX$\to$ZH.

\vspace{-1mm}
\subsection{Hyperparameters}
\vspace{-1mm}
All CLeaD models use AdamW \cite{loshchilov2019adamw} ($\text{lr}{=}10^{-3}$, weight decay $10^{-4}$), temperature $\tau{=}0.1$, loss weighting $\lambda{=}0.5$, and a 128-dimensional projection output with a hidden width of 256; all WavLM weights remain frozen.
Training length differs by experiment for compute reasons and is reported explicitly: the layer-wise ablation (Tables~\ref{tab:monolingual}--\ref{tab:ablation}) uses 100 epochs with batch size 32; the LOSO evaluation (Table~\ref{tab:loso}) uses 5 epochs with batch size 128 per fold, fit on an 80\% speaker subset of each fold's training speakers; the hyperparameter sweep (Table~\ref{tab:hyperparam}) uses 10 epochs with batch size 128; and the projection-space analysis (Sec.~\ref{sec:proj}) uses 30 epochs with batch size 32. Values are therefore comparable within a table but not across tables.
The cold temperature sharpens the contrastive objective, tightening same-class cross-lingual clusters, while equal weighting prevents either term from dominating; a single fixed seed (42) is used throughout, and both choices are examined in Table~\ref{tab:hyperparam}.

\vspace{-1mm}
\subsection{Layer-Wise Ablation Design and Reproducibility}
\vspace{-1mm}
\label{sec:repro}
All six conditions are repeated for WavLM-Base-Plus layers 6--9 and WavLM-Large layers 12, 14, 16 and 18, yielding 240 runs (6 conditions $\times$ 4 layers $\times$ 5 models $\times$ 2 variants), targeting the intermediate depth where prior work finds the most transferable paralinguistic representations, since shallow layers encode local acoustics and deep layers specialize toward language-specific phonetics \cite{maji2024investigation, han2024crosslingual}.
All runs reuse the same speaker-independent splits and fixed E-DAIC subsample; code, loaders, seed 42, split files (\texttt{edaic\_split.csv}, \texttt{modma\_split.csv}) and speaker-ID lists (75/24/23 E-DAIC; 31/11/10 MODMA).
Runs use Python~3.13.7, PyTorch~2.12.0 and scikit-learn~1.8.0; the last matters for Fig.~\ref{fig:tsne}, since t-SNE's \texttt{init}/\texttt{learning\_rate} defaults changed in 1.2 (\texttt{random}/\texttt{200.0} to \texttt{pca}/\texttt{auto}) and \texttt{n\_iter} became \texttt{max\_iter} in 1.5.
We call \texttt{TSNE(n\_components=2, perplexity=30, random\_state=42)} at that version's remaining defaults; both projections ran to the 1000-iteration cap (\texttt{n\_iter\_}~$=999$, zero-indexed), so the \texttt{n\_iter\_without\_progress}$=300$ early stop never triggered. Code is available here: \url{https://github.com/anishxa/wavLM_crosslingual_depressiondet_clead}.

\vspace{-0.5mm}
\section{Experimental Results and Analyses}
\vspace{-1mm}
\subsection{Monolingual Performance}
\vspace{-1mm}
Table~\ref{tab:monolingual} reports Layer 6/12 within-language results. Large significantly outperforms Base-Plus on EN$\to$EN (CLeaD AUC CIs: [0.763,0.787] vs [0.825,0.846], non-overlapping). English LR achieves F1 0.638, vs Chen's 0.74 \cite{chen2025layerwise} under segment-level splits.
Mandarin monolingual performance is substantially lower, reflecting corrected evaluation and use of MODMA (not CMDC).

\begin{table}[!t]
\caption{Monolingual Performance (Base-Plus L6, Large L12). F1 is depressed-class F1. Bold marks the best value in each column within a condition block.}
\vspace{-5mm}
\label{tab:monolingual}
\begin{center}
\scriptsize
\begin{tabular}{llcccc}
\toprule
\textbf{Cond.} & \textbf{Model} & \multicolumn{2}{c}{\textbf{Base-Plus}} & \multicolumn{2}{c}{\textbf{Large}} \\
 & & \textbf{F1} & \textbf{AUC} & \textbf{F1} & \textbf{AUC} \\
\midrule
EN$\to$EN & LR               & 0.638 & 0.785 & 0.644 & 0.796 \\
          & SVM-Linear       & \textbf{0.659} & \textbf{0.796} & 0.679 & 0.823 \\
          & GRU              & 0.588 & 0.667 & 0.375 & 0.632 \\
          & CLeaD            & 0.637 & 0.775 & 0.621 & 0.835 \\
          & CLeaD w/o SupCon & 0.622 & 0.785 & \textbf{0.704} & \textbf{0.866} \\
\midrule
ZH$\to$ZH & LR               & 0.463 & 0.538 & 0.398 & 0.447 \\
          & SVM-Linear       & 0.476 & 0.543 & 0.382 & 0.449 \\
          & GRU              & 0.286 & 0.480 & 0.333 & \textbf{0.640} \\
          & CLeaD            & 0.515 & \textbf{0.585} & \textbf{0.406} & 0.483 \\
          & CLeaD w/o SupCon & \textbf{0.531} & 0.580 & 0.331 & 0.437 \\
\bottomrule
\end{tabular}
\end{center}
\vspace{-7mm}
\end{table}

\vspace{-1mm}
\subsection{Cross-Lingual Transfer}
\vspace{-1mm}
Table~\ref{tab:crosslingual} shows Large at near-zero F1 across all Mandarin conditions, with non-overlapping CIs at every layer pair (e.g., MIX$\to$ZH CLeaD: Base-Plus [0.462,0.513] vs Large [0.184,0.236]), a statistically robust scale reversal.
The Large GRU is degenerate in EN$\to$ZH and MIX$\to$ZH, reaching 5/5 Dep-Rec at 0/5--1/5 HC recall by labelling every speaker depressed, so precision collapses under distribution shift (daggered in the table); in ZH$\to$EN it collapses the other way to F1 0.182, where CLeaD attains the best Base-Plus F1 at 0.505.
CLeaD trails LR/SVM-Linear in MIX$\to$EN (0.491 vs 0.567/0.596) and the GRU in EN$\to$ZH zero-shot (0.332 vs 0.545), but improves MIX$\to$ZH depressed-class recall over its cross-entropy ablation (3/5 at Layer 6, 4/5 at Layers 7--8, versus 2/5 and 1/5); LR and SVM-Linear also reach 3/5 at Layer 6 with better HC recall (5/5 vs 4/5) and LR matches 4/5 at Layer 7, so CLeaD exceeds both only at Layer 8 (Table~\ref{tab:ablation}).
The EN$\to$ZH / ZH$\to$EN asymmetry may reflect acoustic variability between spontaneous E-DAIC and structured MODMA speech, untested here.
\begin{table}[!t]
\caption{Cross-lingual transfer (BP=Base-Plus L6, Lg=Large L12). F1 is depressed-class F1; D/H Rec. gives depressed/healthy recall counts. Bold marks the best value in each column within a condition block, ranking D/H Rec.\ by depressed recall first; ties are bolded jointly.}
\vspace{-5mm}
\label{tab:crosslingual}
\begin{center}
\scriptsize
\setlength{\tabcolsep}{2pt}
\begin{tabular}{llcccc}
\toprule
\textbf{Cond.} & \textbf{Model} & \textbf{BP F1} & \textbf{BP D/H Rec.} & \textbf{Lg F1} & \textbf{Lg D/H Rec.} \\
\midrule
EN$\to$ZH & LR          & 0.213 & 0/5;5/5 & 0.121 & 0/5;5/5 \\
          & SVM-Lin     & 0.236 & 0/5;5/5 & 0.060 & 0/5;5/5 \\
          & GRU         & \textbf{0.545} & \textbf{3/5;2/5} & \textbf{0.714}$^\dagger$ & \textbf{5/5;1/5}$^\dagger$ \\
          & CLeaD       & 0.332 & 1/5;5/5 & 0.028 & 0/5;5/5 \\
          & CLeaD w/o SC & 0.206 & 0/5;5/5 & 0.054 & 0/5;5/5 \\
\midrule
ZH$\to$EN & LR          & 0.408 & -- & \textbf{0.521} & -- \\
          & SVM-Lin     & 0.420 & -- & 0.475 & -- \\
          & GRU         & 0.182 & -- & 0.000 & -- \\
          & CLeaD       & \textbf{0.505} & -- & 0.388 & -- \\
          & CLeaD w/o SC & 0.484 & -- & 0.445 & -- \\
\midrule
MIX$\to$EN & LR         & 0.567 & -- & 0.597 & -- \\
           & SVM-Lin    & \textbf{0.596} & -- & \textbf{0.620} & -- \\
           & GRU        & 0.400 & -- & 0.414 & -- \\
           & CLeaD      & 0.491 & -- & 0.466 & -- \\
           & CLeaD w/o SC & 0.501 & -- & 0.586 & -- \\
\midrule
MIX$\to$ZH & LR         & 0.467 & \textbf{3/5;5/5} & 0.299 & 1/5;4/5 \\
           & SVM-Lin    & 0.486 & \textbf{3/5;5/5} & 0.285 & 1/5;4/5 \\
           & GRU        & 0.333 & 1/5;5/5 & \textbf{0.667}$^\dagger$ & \textbf{5/5;0/5}$^\dagger$ \\
           & CLeaD      & 0.488 & 3/5;4/5 & 0.210 & 0/5;4/5 \\
           & CLeaD w/o SC & \textbf{0.498} & 2/5;4/5 & 0.377 & 2/5;4/5 \\
\bottomrule
\multicolumn{6}{l}{\scriptsize $^\dagger$ Large GRU predicts all speakers as depressed: degenerate, not a useful detector.}
\end{tabular}
\end{center}
\vspace{-7mm}
\end{table}

\vspace{-1mm}
\subsection{Layer Ablation}
\begin{figure}[!b]
\centering
\begin{tikzpicture}
\begin{axis}[
  width=0.9\columnwidth, height=3.2cm,
  xlabel={Layer Pair}, ylabel={F1 (MIX$\to$ZH)},
  xtick={1,2,3,4}, xticklabels={L6/12, L7/14, L8/16, L9/18},
  x tick label style={font=\small},
  ymin=0.15, ymax=0.75,
  legend style={
    at={(0.5,1.05)},
    anchor=south,
    legend columns=3,
    font=\scriptsize,
    cells={anchor=west},
  },
  grid=major, grid style={dashed, gray!30},
  tick label style={font=\small},
  label style={font=\small},
]
\addplot[red, mark=triangle*, thick] coordinates {(1,0.488)(2,0.486)(3,0.561)(4,0.455)};
\addlegendentry{BP CLeaD}
\addplot[orange, mark=o, thick] coordinates {(1,0.498)(2,0.420)(3,0.475)(4,0.528)};
\addlegendentry{BP CLeaD w/o SC}
\addplot[blue, mark=square*, thick] coordinates {(1,0.467)(2,0.531)(3,0.473)(4,0.459)};
\addlegendentry{BP LR}
\addplot[red, mark=triangle*, thick, dashed] coordinates {(1,0.210)(2,0.242)(3,0.370)(4,0.305)};
\addlegendentry{Large CLeaD}
\addplot[blue, mark=square*, thick, dashed] coordinates {(1,0.299)(2,0.318)(3,0.341)(4,0.343)};
\addlegendentry{Large LR}
\end{axis}
\end{tikzpicture}
\vspace{-4mm}
\caption{MIX$\to$ZH F1 across layer pairs. Base-Plus (solid) outperforms Large (dashed) at all layers with non-overlapping 95\% CIs. The Dep-Rec advantage of CLeaD over CLeaD w/o SC peaks at L7 (4/5 vs 1/5, Table~\ref{tab:ablation}), whereas peak segment F1 occurs at L8 (0.561).}
\label{fig:ablation}
\end{figure}
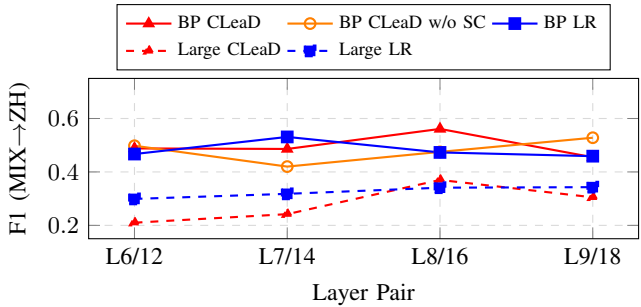

\vspace{-1mm}
Fig.~\ref{fig:ablation} and Table~\ref{tab:ablation} show MIX$\to$ZH results across layers, now including SVM-Linear for comparison with the strongest linear baseline.
Base-Plus CLeaD peaks at Layer 8 (F1 0.561, Dep-Rec 4/5).
The SupCon gap is most pronounced at Layer 7, where CLeaD reaches Dep-Rec/HC-Rec of 4/5 and 4/5 against 1/5 and 5/5 for CLeaD w/o SupCon and 2/5 and 4/5 for SVM-Linear, though LR also reaches 4/5 there with 5/5 HC-Rec; Layer 8 (4/5 and 3/5) is the only layer where CLeaD exceeds both linear baselines on depressed-class recall.
Contrastive alignment therefore adds sensitivity relative to its cross-entropy counterpart, while its advantage over tuned linear baselines is confined to a single layer; at $n{=}5$ these comparisons are indicative rather than conclusive.
Large consistently underperforms Base-Plus at all layers with non-overlapping CIs.
Two further patterns stand out: the GRU attains the highest segment F1 and AUC at most layers (AUC 0.8 at Layers 6 and 7) yet yields only 1/5 and 2/5 Dep-Rec, suggesting useful ranked probabilities but poor calibration for speaker-level decisions; and at Layer 9, CLeaD w/o SupCon (4/5) matches the GRU and outperforms CLeaD (2/5).
We lack a definitive explanation for this single-point reversal, but it reinforces deploying CLeaD at Layers 7--8 rather than Layer 9.

\begin{table}[!t]
\caption{Layer Ablation: MIX$\to$ZH Base-Plus. F1 is depressed-class F1. Bold marks the best value in each column within a layer block; ties are bolded jointly.}
\vspace{-5mm}
\label{tab:ablation}
\begin{center}
\footnotesize
\begin{tabular}{clccc}
\toprule
\textbf{L} & \textbf{Model} & \textbf{F1} & \textbf{AUC} & \textbf{Dep-Rec} \\
\midrule
\multirow{5}{*}{6} & LR               & 0.467 & 0.503 & \textbf{3/5} \\
                   & SVM-Linear       & 0.486 & 0.515 & \textbf{3/5} \\
                   & GRU              & 0.333 & \textbf{0.800} & 1/5 \\
                   & CLeaD            & 0.488 & 0.521 & \textbf{3/5} \\
                   & CLeaD w/o SupCon & \textbf{0.498} & 0.555 & 2/5 \\
\midrule
\multirow{5}{*}{7} & LR               & 0.531 & 0.544 & \textbf{4/5} \\
                   & SVM-Linear       & 0.470 & 0.472 & 2/5 \\
                   & GRU              & \textbf{0.571} & \textbf{0.800} & 2/5 \\
                   & CLeaD            & 0.486 & 0.539 & \textbf{4/5} \\
                   & CLeaD w/o SupCon & 0.420 & 0.526 & 1/5 \\
\midrule
\multirow{5}{*}{8} & LR               & 0.473 & 0.501 & 3/5 \\
                   & SVM-Linear       & 0.482 & 0.492 & 3/5 \\
                   & GRU              & \textbf{0.571} & \textbf{0.600} & 2/5 \\
                   & CLeaD            & 0.561 & 0.517 & \textbf{4/5} \\
                   & CLeaD w/o SupCon & 0.475 & 0.521 & 3/5 \\
\midrule
\multirow{5}{*}{9} & LR               & 0.459 & 0.492 & 2/5 \\
                   & SVM-Linear       & 0.443 & 0.471 & 2/5 \\
                   & GRU              & \textbf{0.727} & \textbf{0.720} & \textbf{4/5} \\
                   & CLeaD            & 0.455 & 0.505 & 2/5 \\
                   & CLeaD w/o SupCon & 0.528 & 0.515 & \textbf{4/5} \\
\bottomrule
\end{tabular}
\end{center}
\vspace{-7mm}
\end{table}

\vspace{-1.5mm}
\subsection{LOSO Evaluation (Primary Mandarin Metric)}
Table~\ref{tab:loso} reports LOSO on all 52 MODMA speakers.
SVM-Linear leads with Spk~F1 0.762 and LR achieves 0.714, reflecting the advantage of a tuned linear boundary at this dataset size; CLeaD (0.640) outperforms CLeaD w/o SupCon (0.622) at Base-Plus Layer 7, a positive speaker-level contribution beyond the MLP architecture alone, though the 0.018 margin is small and reported without a per-comparison confidence interval (see Sec.~\ref{sec:conclusion}).
CLeaD is therefore not the top model by Spk~F1, trailing both SVM-Linear and LR at Base-Plus L7; its contribution is analyzing when SupCon helps rather than claiming state of the art, and it yields a consistent, modest improvement in depressed-class recall, matching our clinical motivation.

\begin{table}[!b]
\caption{LOSO evaluation on MODMA (primary Mandarin metric, 52 speakers). Spk~F1 is speaker-level depressed-class F1 after majority voting; AUC is reported at the segment level, since speaker-level decisions are majority votes rather than calibrated scores. Bold marks the best value in each column within a variant block.}
\vspace{-5mm}
\label{tab:loso}
\begin{center}
\scriptsize
\begin{tabular}{llcc}
\toprule
\textbf{Variant} & \textbf{Model} & \textbf{Spk F1} & \textbf{Seg AUC} \\
\midrule
Base-Plus L7 & LR               & 0.714 & 0.703 \\
             & SVM-Linear       & \textbf{0.762} & \textbf{0.704} \\
             & CLeaD            & 0.640 & 0.654 \\
             & CLeaD w/o SupCon & 0.622 & 0.690 \\
\midrule
Large L14    & LR               & 0.667 & \textbf{0.707} \\
             & SVM-Linear       & 0.650 & 0.704 \\
             & CLeaD            & \textbf{0.691} & 0.686 \\
             & CLeaD w/o SupCon & 0.667 & 0.647 \\
\bottomrule
\end{tabular}
\end{center}
\end{table}

\vspace{-1mm}
\subsection{Hyperparameter Sensitivity}
Table~\ref{tab:hyperparam} supports $\tau{=}0.1$, $\lambda{=}0.5$ at Layer 7 (the primary MIX$\to$ZH layer).
Colder temperatures ($\tau{=}0.05$) cap peak Dep-Rec at 3/5, and strong down-weighting of CE ($\lambda{=}0.7$) does the same at every temperature, indicating that neither alignment nor classification alone is sufficient.
Warmer temperatures are not uniformly worse: $\tau{=}0.2$ with $\lambda{=}0.3$ also reaches 4/5 and attains a higher segment F1 (0.542), so the chosen setting is one of two comparable optima rather than a clear winner.
The sweep uses the shorter 10-epoch, batch-128 schedule noted in Sec.~\ref{sec:repro}, so its absolute F1 at the chosen setting (0.522) is not directly comparable with the 100-epoch Layer-7 entry in Table~\ref{tab:ablation} (0.486); the sweep ranks configurations rather than reproducing Table~\ref{tab:ablation}.

\begin{table}[!t]
\caption{Hyperparameter Sensitivity: CLeaD MIX$\to$ZH Base-Plus L7.
The bolded row is the chosen configuration ($\tau=0.1, \lambda=0.5$), not the column maximum.}
\vspace{-5mm}
\label{tab:hyperparam}
\begin{center}
\begin{tabular}{ccccc}
\toprule
$\tau$ & $\lambda$ & \textbf{F1} & \textbf{AUC} & \textbf{Dep-Rec} \\
\midrule
0.05 & 0.3 & 0.465 & 0.506 & 2/5 \\
0.05 & 0.5 & 0.370 & 0.479 & 1/5 \\
0.05 & 0.7 & 0.476 & 0.474 & 3/5 \\
0.10 & 0.3 & 0.482 & 0.516 & 3/5 \\
\textbf{0.10} & \textbf{0.5} & \textbf{0.522} & \textbf{0.506} & \textbf{4/5} \\
0.10 & 0.7 & 0.487 & 0.496 & 3/5 \\
0.20 & 0.3 & 0.542 & 0.508 & 4/5 \\
0.20 & 0.5 & 0.442 & 0.514 & 2/5 \\
0.20 & 0.7 & 0.471 & 0.496 & 3/5 \\
\bottomrule
\end{tabular}
\end{center}
\vspace{-7mm}
\end{table}

\vspace{-1mm}
\subsection{Projection Space Analysis}
\begin{figure}[!t]
\centering
\includegraphics[width=0.95\columnwidth]{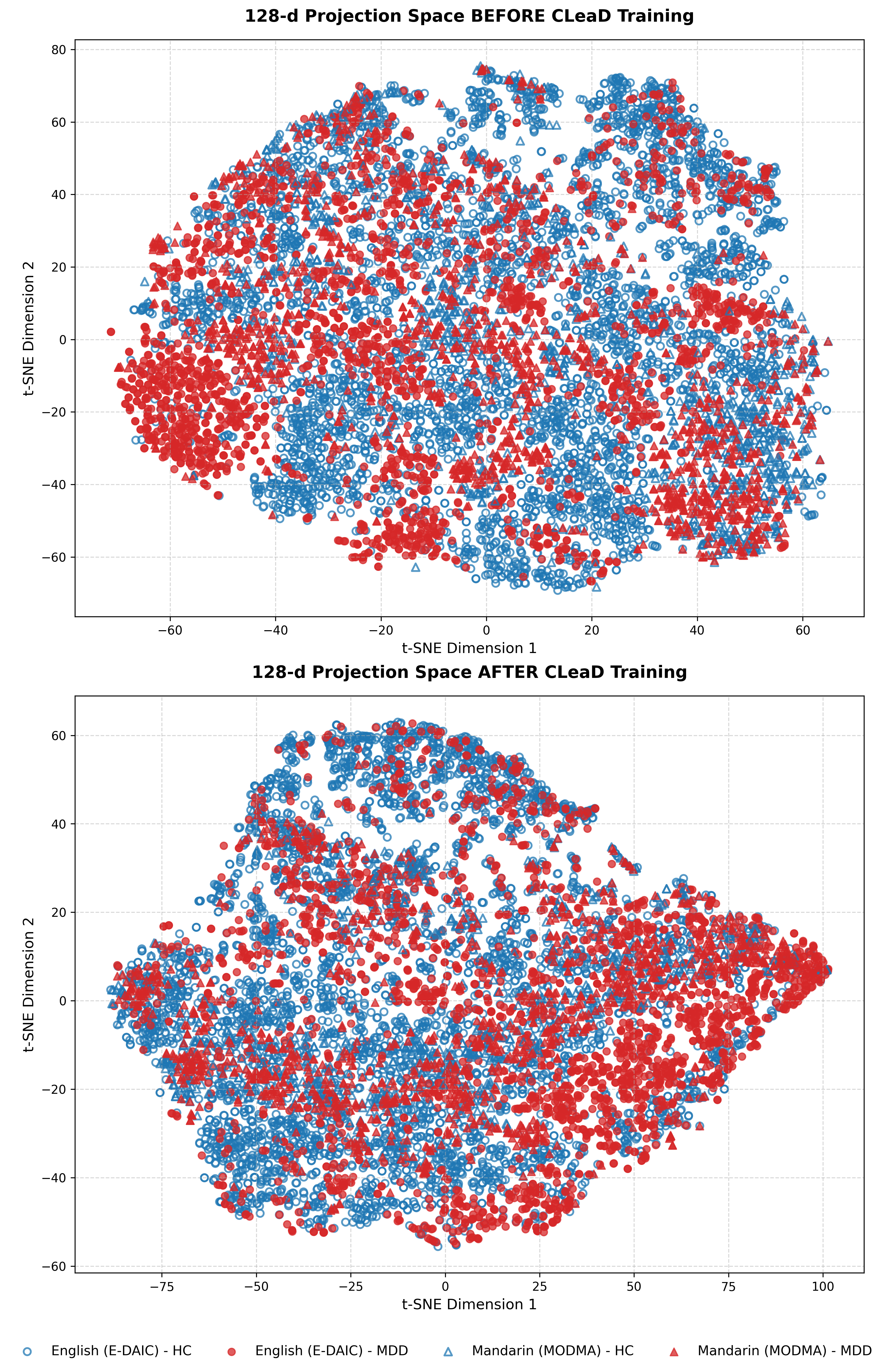}
\caption{Layer-7 t-SNE before (top) and after (bottom) MIX$\to$ZH training. Circles/triangles denote English/Mandarin; filled/open markers denote MDD/HC.}
\label{fig:tsne}
\vspace{-6mm}
\end{figure}

\vspace{-1mm}
\label{sec:proj}
Fig.~\ref{fig:tsne} shows Layer-7 t-SNE (perplexity 30, seed 42) over the held-out E-DAIC and MODMA test segments (6{,}040 English, 2{,}394 Mandarin), where \emph{before} is the randomly initialized projection head and \emph{after} is the same head trained for 30 epochs on MIX. Metrics use the normalized 128-D embeddings, not the 2-D coordinates. A language classifier (logistic regression at library defaults, stratified 5-fold) falls from 89.2\% to 84.0\% accuracy; because the pool is 71.6\% English, the majority-class baseline is 71.6\% rather than 50\%, so the drop indicates reduced but not eliminated language separability. Clinical and language silhouettes move from 0.010 to 0.066 and from 0.020 to 0.018; being near zero, these are directional support only, and the classifier is the stronger evidence.

\vspace{-1mm}
\section{Discussion}
\vspace{-1mm}
\subsection{Why Model Scale Hurts Cross-Lingual Transfer}
\vspace{-1mm}
\label{sec:scale}
WavLM-Large improves monolingual English, consistent with SUPERB \cite{yang2021superb}, yet underperforms Base-Plus on cross-lingual Mandarin at every paired layer (non-overlapping CIs). A plausible cause is stronger English-specific phonetic/prosodic specialization after 94K hours of English pretraining; related multilingual-ASR and feature-entanglement findings support, but do not establish, this mechanism \cite{shi2023mlsuperb,zhangEntangle}.

\vspace{-1mm}
\subsection{Why Intermediate Layers Transfer Best}
\vspace{-1mm}
Base-Plus Layers 6--8 transfer best, consistent with intermediate prosodic/temporal cues being less site-specific than shallow acoustics and less English-specialized than deep features \cite{kappen2023speech,han2024crosslingual}. CLeaD peaks at Layer 8 (F1 0.561) and reaches 4/5 Dep-Rec at Layers 7--8; Layer 6 remains the cross-paper anchor \cite{maji2024investigation}.

\vspace{-1mm}
\subsection{Zero-Shot Failure and Future Directions}
\vspace{-1mm}
In EN$\to$ZH zero-shot, CLeaD detects only 1/5 depressed speakers versus the GRU's 3/5. Without Mandarin samples, the cross-lingual SupCon gradient is absent, reducing CLeaD to standard supervised contrastive learning. Target-language pseudo-labels or memory banks, and multilingual backbones such as XLSR-53 or MMS, are natural remedies.

\vspace{-1mm}
\subsection{Clinical Implications}
\vspace{-1mm}
CLeaD targets first-stage multilingual screening, where a missed patient is costlier than a false positive reviewed by a clinician; we therefore report depressed-class recall with LOSO Spk~F1. Three findings emerge: mixed-language training raises Mandarin Dep-Rec from 1/5 to 3/5--4/5; Base-Plus transfers better than Large; and the leakage ablation shows why speaker-independent evaluation is essential.

\vspace{-1mm}
\section{Conclusion}
\vspace{-1mm}
\label{sec:conclusion}
Under LOSO evaluation on 52 MODMA speakers, CLeaD yields a modest gain over its cross-entropy counterpart (Spk~F1 0.640 vs. 0.622) and improves depressed-class recall at Base-Plus Layers 7--8 in the MIX$\to$ZH condition. Given the small test set, single seed, and stronger linear baselines, this localizes where SupCon helps rather than establishing state of the art. WavLM-Large degrades cross-lingual performance despite stronger English results, while speaker leakage inflates Mandarin F1 by 0.23. Corpus confounds remain; multilingual backbones, multi-seed runs, and adaptation baselines are priorities. Code, splits, and complete configurations are released via Sec.~\ref{sec:repro}.

\vspace{-1mm}
\section*{Acknowledgment}
\vspace{-1mm}
This work was supported by the U.S. National Science Foundation through ACCESS allocation CIS260814 and CIS260813; the National Science and Technology Council, Taiwan, under Grant 114-2917-I-564-030 awarded to Huang-Cheng Chou; and, in part, the Advanced Research Projects Agency for Health (ARPA-H). The views and conclusions should not be interpreted as representing official U.S. Government policies, either expressed or implied.

\bibliographystyle{IEEEtran}
\bibliography{IEEEexample}

\end{document}